# Widely tunable and narrow linewidth chip-scale lasers from deep visible to near-IR

Mateus Corato-Zanarella[1], Andres Gil-Molina[1], Xingchen Ji[1], Min Chul Shin[1], Aseema Mohanty[2], Michal Lipson[1]

[1]Department of Electrical Engineering, Columbia University, New York, New York 10027, USA

[2]Department of Electrical and Computer Engineering, Tufts University, Medford, MA 02155, USA

## Abstract

Widely-tunable and narrow-linewidth integrated lasers across all visible wavelengths are necessary to enable on-chip technologies such as quantum photonics, optical trapping, and biophotonics. However, such lasers have not been realized due to the strong wavelength sensitivity and high loss of integrated optical devices at short wavelengths. Here we overcome these challenges and demonstrate a chip-scale visible lasers platform by using tightly-confined, micrometer-scale silicon nitride resonators and commercial Fabry-Perot laser diodes. We achieve tunable and narrow-linewidth lasing in the deep blue (450 nm), blue (488 nm), green (520 nm), red (660 nm) and near-IR (785 nm) with coarse tuning up to 12 nm, fine tuning up to 33.9 GHz, linewidth down to sub-kHz, side-mode suppression ratio > 35 dB, and fiber-coupled power up to 10 mW. These specifications of our chip-scale lasers are comparable to those of state-of-the-art commercial laser systems, making them stand out as powerful tools for the next generation of visible-light technologies.

# Main

On-chip widely tunable and narrow linewidth lasers across all visible wavelengths are necessary to enable chip-scale technologies such as quantum photonics[1,2], optical trapping[1], AR/VR[3], and biophotonics[4,5]. The realization of such lasers, however, has been prevented by the strong wavelength sensitivity and high loss of integrated optical devices at short wavelengths. There are currently two types of visible lasers, neither of which addresses the need for an on-chip tunable and narrow linewidth laser platform for the whole visible spectrum. One state-of-the-art laser type is the Fabry-Perot (FP) laser diodes. These lasers are compact and inexpensive but not single frequency and not easily tunable, making them unsuitable for many important applications such as quantum photonics and spectroscopy. The other state-of-the-art laser type is the tunable and narrow linewidth lasers that use free-space components and external cavities[6–9], making them bulky, expensive, and not suitable for chip-scale integration. The ideal architecture would be a high-performance integrated visible lasers platform that bridges the gap between these two technological paradigms. This high-performance, on-chip visible lasers platform must be compact, inexpensive, provide wide tunability and narrow linewidth, and cover all colors of light. These novel lasers are critical not only to facilitate and speed up research, but also to deploy resulting technologies for use in the real world. In order to realize such lasers, however, a new integrated platform that overcomes the strong wavelength sensitivity and high loss of optical devices at visible wavelengths is required.

Here we demonstrate a chip-scale visible lasers platform that address these issues and allows for wide tunability, narrow linewidth, and robust lasing across all visible wavelengths from deep blue (450 nm) to near-IR (785 nm) (Fig 1a). Specifically, our single-chip platform uses tightly-confined, micrometer-scale, high-quality factor (Q) silicon nitride ($Si_3N_4$) resonators and commercial Fabry-Perot (FP) laser diodes (Fig 1b). Our unique resonator design and the use of FP laser diodes as light sources are the two key aspects that allow us to achieve broadband tunability, narrow linewidth, and robustness. First, tunability requires tunable, frequency-selective optical feedback and is traditionally limited by the narrow optical bandwidth of the components that provide the feedback. Our design leverages the high confinement of $Si_3N_4$ to make micrometer-scale resonators with large free spectral range (FSR) for all visible wavelengths, an approach that translates into wide mode-hop free tunability with high side-mode suppression ratio (SMSR). Second, narrow linewidth requires low loss, a characteristic hindered by the dramatic increase of surface scattering loss at short wavelengths[10]. Our platform uses a novel resonator structure that decreases loss by minimizing the overlap between optical mode and sidewalls without sacrificing broadband operation across the entire visible spectrum. Our strategy enables the generation of narrow-linewidth and highly coherent light of all colors, a necessary condition for critical new technologies in quantum optics, optical cooling and trapping, and atomic clocks. Third, stable lasing requires robustness to coupling loss and unwanted reflections, two hindrances that are intrinsically stronger at deeper visible wavelengths and cannot be fully eliminated, even with optimal packaging. To address this challenge, our design uses FP laser diodes as the light sources to be self-injection locked to the high-Q resonators with high tolerance against coupling loss and unwanted reflections[11]. This specific design element makes it possible to achieve high-performance lasers of all colors within the same chip unlike all other state-of-the-art laser platforms. By using broadband ring resonator-based feedback loops to self-injection lock FP laser diodes (Fig. 1b) we achieve tunable, narrow linewidth, and robust near-octave lasing from deep visible (450 nm) to near-IR (785 nm) in a single chip.

The broadband and low-loss optical feedback scheme based on ring resonators is a key design feature that enables tunable and narrow linewidth lasing of all light colors. First, the design ensures mode-hop free tuning by making the ring small (10 µm bending radius) (Fig. 1a, inset). Due to its large FSR of several nanometers for all visible wavelengths (supplementary information), the resonators provide selective feedback to a single longitudinal mode of the FP laser diodes. In contrast to using Vernier filters[12,13], our approach requires fewer phase shifters and thus eliminates unnecessary noise and thermal crosstalk, reduces power consumption, and keeps the footprint small. Second, broadband narrow linewidth is achieved by operating the high Q resonators in single-mode regime with strong coupling to the bus waveguides.

These two requirements are satisfied simultaneously by tapering the ring width from 1500 nm to 300 nm (Fig. 1a, inset)[14]. The highly multimode sections (1500 nm) decrease loss by reducing the overlap between the optical mode and the waveguide sidewalls[15]. This overlap reduction is critical for broadband low loss because surface scattering increases drastically at visible wavelengths[10]. By slowly tapering the width at the coupling regions down to 300 nm, only the fundamental mode is excited and propagated while retaining low loss (Fig. 1c). This width narrowing also results in broadband strong coupling by increasing the modal overlaps between the waveguides and the ring, ensuring that enough power is dropped to the feedback loop and is reflected back to the FP laser diodes to cause self-injection locking. Our novel ring resonator design allows for strong and tunable optical feedback with low loss across all visible wavelengths, a design feature necessary to enable wide tunability and narrow linewidth for lasers of all colors of light.

The choice of FP laser diodes as light sources to leverage their robust self-injection locking against coupling loss and unwanted reflections[11] is another unique aspect of our platform. This robustness stems from the fact that the interface between the FP laser diode and the photonic chip is not part of the main laser cavity formed by the FP laser facets. Since lasing modes already exist, single-frequency lasing is achieved by providing enough optical feedback to a desired mode in order to collapse all optical power to it through self-injection locking, regardless of the injection current[11,16–18]. This robustness contrasts with the sensitivity of the traditional approach that uses reflective semiconductor optical amplifiers (RSOA) as light sources, where the interface between the RSOA and the photonic chip is part of the main laser cavity. In the RSOA approach, the resulting sensitivity to coupling loss and unwanted reflections restricts RSOA-based lasers to the red wavelength range (685 nm) of the visible spectrum and to low injection currents despite optimal packaging and integrated components[12]. We resolve this sensitivity problem through the unique robustness and scalability of our FP lasers-based platform to enable stable, tunable and narrow linewidth lasing from deep visible to near-IR in the same photonic chip.

We collapse all the longitudinal modes of the edge-coupled FP laser diodes (Fig. 2a) into single-frequency lines via independent phase control of the bus waveguide and ring resonator, turning low-cost FP laser sources into narrow linewidth and tunable lasers with high performance previously seen only in expensive and bulky commercial laser systems. We use the ring phase shifter to control the wavelength of the reflected light, and the bus phase shifter to adjust its phase. With the resonator detuned, the power dropped to the feedback loop is negligible (Fig. 2b, top) and the FP diode lases with multiple longitudinal modes (Fig. 2b, bottom). By aligning the ring resonance to a mode of the FP laser, we drop optical power to the feedback loop and reflect it back to the diode (Fig. 2c, top). By further adjusting the phase of the reflected light, it constructively interferes with the light inside the FP laser to act as an additional effective gain for a chosen wavelength. This feedback causes self-injection locking and converts the output of the chip into a single-frequency laser with narrow linewidth (Fig. 2c, bottom). We coarsely tune the wavelength by aligning the ring resonance to different longitudinal modes of the FP laser. We further fine tune the wavelength in between FP modes via frequency pulling after locking to the desired mode. The frequency pulling is achieved by keeping the bus phase shifter fixed and tuning the ring phase shifter to move its resonance. The ring resonance pulls the optical feedback with it, resulting in a change in the lasing wavelength. This self-injection locking process using integrated phase shifters allows wavelength tuning without moving parts or mechanical actuators.

## Results

We demonstrate tunable lasing in the deep blue (450nm), blue (488 nm), green (520 nm), red (660 nm) and near-IR (785 nm) wavelength ranges with coarse tuning up to 12 nm, fine tuning up to 33.9 GHz, side-mode suppression ratio (SMSR) > 35 dB and fiber-coupled power up to 10 mW. For the characterization of our chip-scale lasers, the chip output is collected using edge-coupled cleaved fibers (Fig. 2a and supplementary information). Coarse tuning is measured from deep blue to near-IR by overlapping the lasers output spectra acquired using an optical spectrum analyzer (OSA) (Fig. 3a). The coarse tuning ranges and peak fiber-coupled powers for the different colors are 4.49 nm, 1.74 mW for deep blue; 5.62 nm, 1.75 mW for blue; 4.47 nm, 9.89 mW for green; 4.24 nm, 4.42 mW for red; and 12.00 nm, 10.00 mW for near-IR (Table 1). The tuning ranges are limited only by the gain bandwidth of the FP laser diodes. We achieve fine tuning ranges of 2.0 GHz in deep blue, 3.7 GHz in blue, 8.2 GHz in green, 11.3 GHz in red, and 33.9 GHz in near-IR (Fig. 3b). Fine tuning ranges are measured by changing the power in the ring microheater while monitoring the lasing wavelength with a wavemeter after locking the lasers. Our integrated visible lasers possess tunability and output power levels comparable to state-of-the-art commercial laser systems.

We measure the intrinsic linewidths at the two extremes of the spectrum, blue and near-IR, and obtain ≤ (8 ± 2) kHz and ≤ (0.6 ± 0.2) kHz respectively, both values limited by our instruments. We determine the linewidth in blue by measuring the RF beat note between our integrated laser and a commercial narrow-linewidth laser using a spectrum analyzer. The beat note represents the measured lineshape and is limited by the lineshape of the commercial laser. By fitting the beat note with a Voigt profile, we extract both the Lorentzian contribution, which corresponds to the white noise that defines the intrinsic linewidth, as well as the Gaussian contribution, which corresponds to the flicker and technical noises that broaden the effective linewidth[19]. The Voigt fitting of the lineshape (Fig. 4a) gives a Lorentzian linewidth of (8 ± 2) kHz and a Gaussian linewidth of (250 ± 20) kHz. Since the commercial laser has an intrinsic linewidth on the order of 10 kHz and an effective linewidth between 100 kHz and 500 kHz, we conclude that our measurement is limited by its linewidth. We measure the lineshape and frequency noise of our integrated near-IR laser with a linewidth analyzer. The Voigt fittings of the lineshape (Fig. 4b) for different measurements give an intrinsic linewidth of (0.6 ± 0.2) kHz, limited by the linewidth analyzer sensitivity. The frequency noise gives information about all noise sources that cause the lasing wavelength to fluctuate (Fig.4c). The low frequency components are due to slow variations caused by technical noise such as mechanical vibrations and thermal fluctuations that cause a broadening of the effective linewidth of the laser and correspond to the Gaussian linewidth, while the intrinsic linewidth is determined only by the high frequency components ($\pi$ times the value of the red dashed line in Fig. 4c)[20]. The frequency noise of our laser is below the sensitivity of the instrument, whose noise floor is in between 100 Hz$^2$/Hz and 400 Hz$^2$/Hz. This noise floor places an upper limit for the intrinsic linewidth between 314 Hz and 1.26 kHz. Using the measured and estimated parameters of our system, we estimate that the intrinsic linewidths of our lasers across the whole visible spectrum should be in the range from 0.2 kHz to 5 kHz (Table 1 and supplementary information). The narrow linewidth estimates are supported by the high SMSRs (higher than 35 dB) measured for all lasers, implying that possible linewidth broadening contributions caused by mode partition noise and nonlinear gain are negligible[21]. Our demonstrated linewidths are comparable to state-of-the-art narrow-linewidth commercial lasers and meet coherence requirements of most visible light applications.

## Discussion

Our novel chip-scale visible lasers exhibit the key specifications of tuning range, linewidth, power, and SMSR comparable to bulky commercial laser systems, making them stand out as powerful tools for the next generation of visible-light technologies. Our platform does not require free-space optical components, has a micrometer-scale footprint, and works robustly across almost an octave. Due to its small footprint and broadband operation, our platform is easily scalable and enables multiple lasers of different wavelengths to be integrated on the same chip. Such compact and highly coherent multi-color laser engine is the ideal source for quantum photonics, biophotonics, AR/VR, atomic clocks, and spectroscopy. The self-injection locking robustness also allows for scaling to unprecedented on-chip power levels by using commercial high-power (> 1W) visible laser diodes, opening the door to fully integrated high-power sources for nonlinear optics and optical cooling and trapping. Because our $Si_3N_4$ platform is annealing-free and crack-free, wafer-scale, foundry-compatible manufacturing is possible, a critical technical detail for enabling inexpensive mass production and large-scale deployment. These unique features of our chip-scale platform break the existing paradigm that high-performance visible lasers need to be large and expensive, paving the way for their deployment in fully integrated visible-light systems for applications in the life sciences, atomic research, and vision sciences.

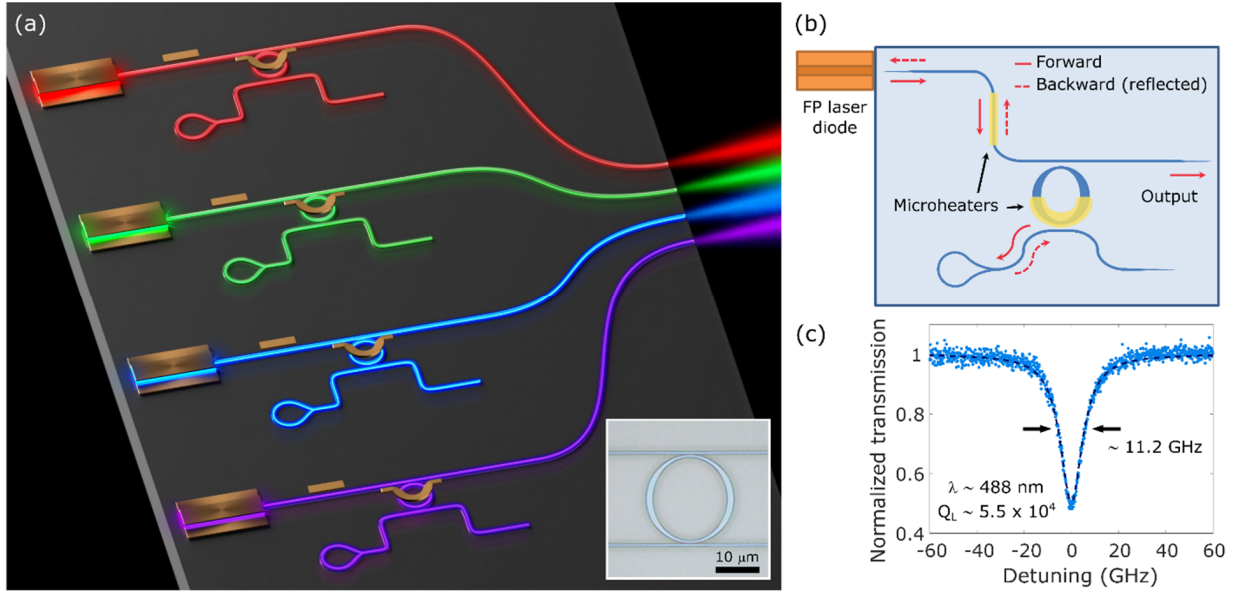

**Figure 1: Chip-scale, multi-wavelength visible lasers.** (a) Artistic view of our integrated platform. A single chip generates narrow linewidth and tunable visible light covering all colors. Inset: Microscope image of the ring resonator before the microheater deposition. Waveguides are made of 175 nm-thick silicon nitride ($Si_3N_4$) surrounded by silicon oxide (see Methods for fabrication details). The ring has a bending radius of 10 µm and a width that tapers along its circumference. The width is 300 nm at the coupling regions and slowly widens to 1500 nm in between. The large FSR prevents mode-hopping, and the tapering ensures single-mode operation while maintaining high Q despite the small footprint. (b) Schematic of a single integrated visible laser using our platform. A Fabry-Perot (FP) laser diode emitting light of the desired color is edge-coupled to the $Si_3N_4$ chip. A ring resonator with a feedback loop at the drop port reflects part of the resonant light back to the FP laser, which leads to self-injection locking. The selection of the lasing wavelength is done via phase shifters (microheaters) on top of the ring resonator and the bus waveguide. (c) Example of ring resonance measured in blue (wavelength around 488 nm). The loaded

quality factor is $5.5 \times 10^4$ with an extinction ratio of 50 %. Resonance measurements at other wavelengths are provided in the supplementary information.

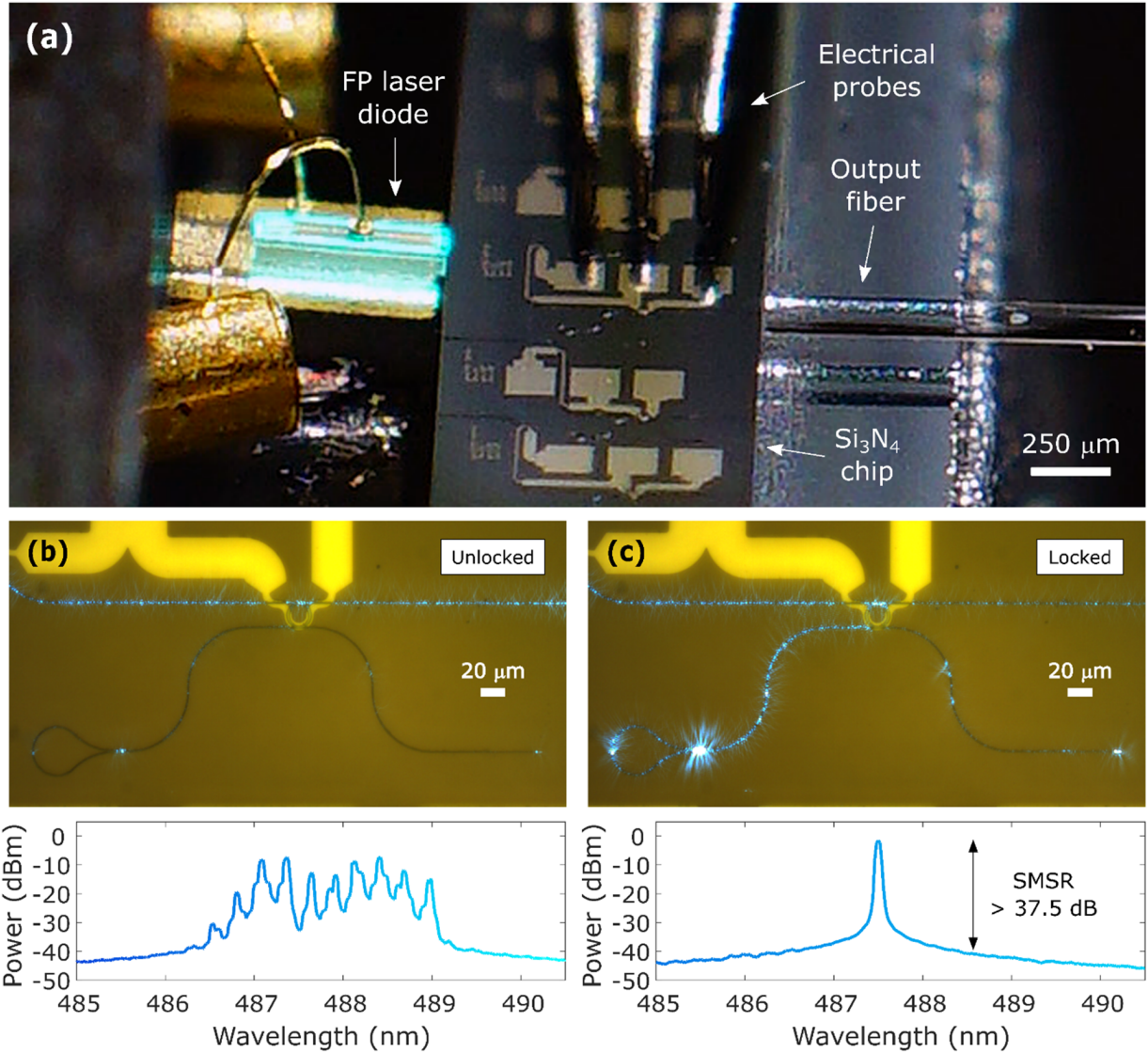

**Figure 2: Example of chip-scale laser (blue wavelength range) and its operation.** (a) Photo of the integrated blue laser. The FP laser diode is edge-coupled to the $Si_3N_4$ chip and the output is collected with a cleaved fiber. The thermo-optic phase shifters are driven by electrical probes. (b) Ring resonance is detuned from the Fabry-Perot (FP) laser diode modes, resulting in no light in the feedback loop (top). The chip output resembles the usual output of the FP laser with multiple lasing modes (bottom). (c) Ring resonance is tuned to one of the FP modes with the feedback loop reflecting part of the light back to the diode (top). Self-injection locking causes all the longitudinal modes to collapse into a single frequency, narrow linewidth lasing mode with high (> 37.5 dB) side mode suppression ratio (SMSR) (bottom). The optical spectra are measured with an optical spectrum analyzer (Ando AQ6314A) and images for the other wavelengths are provided in the supplementary information.

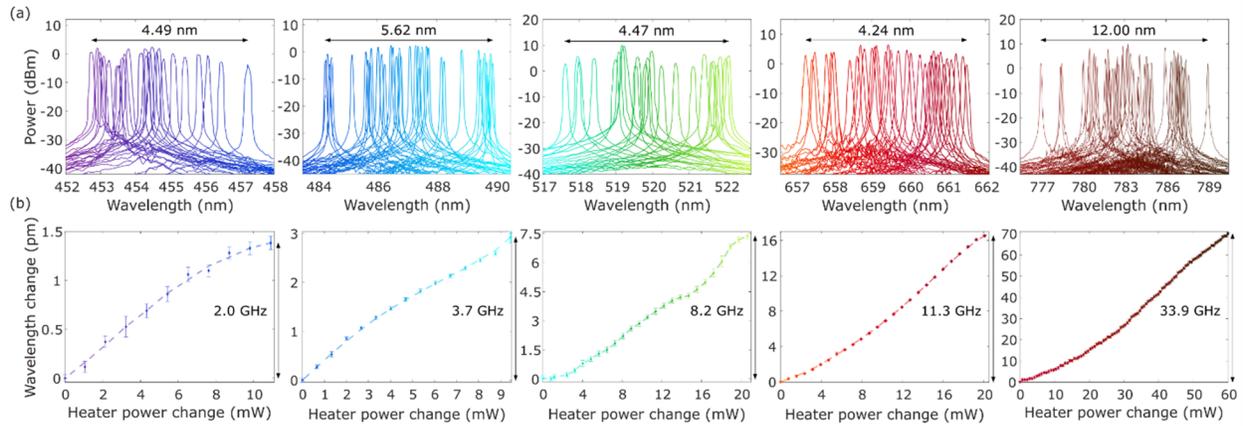

**Figure 3: Integrated lasers coarse and fine wavelength tuning from deep blue to near-IR.** (a) Coarse tuning ranges at deep blue, blue, green, red and near-IR. Optical spectra are measured with an optical spectrum analyzer (Ando AQ6314A) and overlaid to show the tuning. (b) Fine tuning (frequency pulling) as a function of the power applied to the ring microheater at the same wavelength ranges. Detuning from the initial wavelength is measured with a wavemeter (Advantest Q8326 for the near-IR and HighFinesse WSU-10 for the other wavelengths). See supplementary information for detailed characteristics of the Fabry-Perot laser diodes.

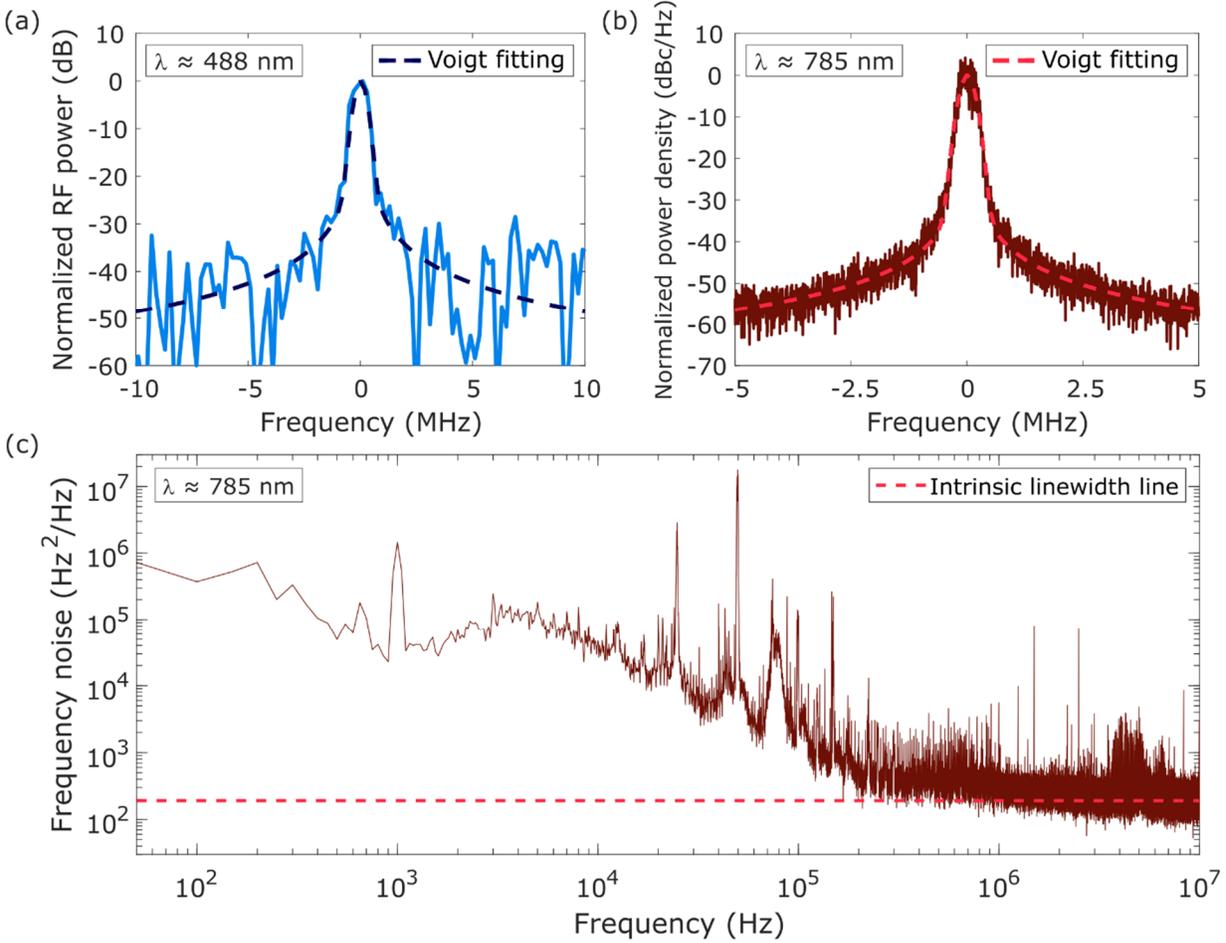

**Figure 4: Linewidth characteristics of the blue and near-IR integrated lasers.** (a) RF beat note between our integrated blue laser and a commercial narrow-linewidth blue laser (Toptica DL Pro 488 nm). The beat note represents the measured lineshape and is limited by the lineshape of the commercial laser. By fitting the beat note with a Voigt profile, we extract the Lorentzian contribution, which corresponds to the white noise that defines the intrinsic linewidth, as well as the Gaussian contribution, which corresponds to the flicker and technical noises that broaden the effective linewidth. The Voigt fitting results in an intrinsic (Lorentzian) linewidth of $(8 \pm 2)$ kHz and a Gaussian linewidth of $(250 \pm 20)$ kHz. Since the commercial laser has an intrinsic linewidth on the order of 10 kHz and an effective linewidth between 100 kHz and 500 kHz, we conclude that our measurement is limited by its linewidth. We measured the beat note using a spectrum analyzer (Agilent E4407B) with a sweep time of 4.990 ms and a resolution bandwidth of 30 kHz.

(b) Example lineshape of our integrated near-IR laser. Voigt fittings of several measurements give an intrinsic linewidth of (0.6 ± 0.2) kHz, limited by our instrument. This example measurement has an intrinsic linewidth of (0.6 ± 0.1) kHz and a Gaussian linewidth of (143 ± 3) kHz. (c) Frequency noise of our integrated near-IR laser corresponding to the lineshape in (b). The red dashed line corresponds to the high frequency components that determine the intrinsic linewidth. We measured both (b) and (c) with a linewidth analyzer (HighFinesse LWA-1k 780). The frequency noise of our laser is below the sensitivity of the instrument, whose noise floor is in between 100 Hz$^2$/Hz and 400 Hz$^2$/Hz. This noise floor places an upper limit for the intrinsic linewidth between 314 Hz and 1.26 kHz.

**Table 1: Summary of the specifications of our chip-scale visible lasers from deep blue to near-IR.**

| Wavelength (nm) | Coarse tuning (nm) | Fine tuning (GHz) | Measured linewidth (kHz) | Estimated linewidth (kHz) | SMSR (dB) | Fiber-coupled power (mW) |
|---|---|---|---|---|---|---|
| 785 | 12.00 | 33.9 | ≤ (0.6 ± 0.2) * | < 0.9 | ≥ 35 | 10.00 |
| 660 | 4.24 | 11.3 | - | < 5.0 | ≥ 35 | 4.42 |
| 520 | 4.47 | 8.2 | - | < 0.2 | ≥ 35 | 9.89 |
| 488 | 5.62 | 3.7 | ≤ (8 ± 2) * | < 1.2 | ≥ 35 | 1.75 |
| 450 | 4.49 | 2.0 | - | < 3.4 | ≥ 35 | 1.74 |

* Limited by measurement instruments

# Methods

## Fabrication methods

To fabricate the silicon nitride ($Si_3N_4$) photonic chip, we first grow 1 µm of wet thermal silicon oxide ($SiO_2$) on a silicon wafer and deposit 175 nm of low-pressure chemical vapor deposition (LPCVD) $Si_3N_4$. Next, we pattern the waveguides and resonators using electron-beam lithography and fluorine-based etch. We clad the devices with 1 µm of plasma-enhanced chemical vapor deposition (PECVD) $SiO_2$. For the microheaters, we use photolithography and a metal lift-off process to pattern 100 nm of platinum on a titanium adhesion layer. Finally, we define the facets of the photonic chip by etching 150 µm into the silicon substrate using a Bosch etch process and dice the wafer to separate the individual chip.

## Experimental setup

The Fabry-Perot (FP) laser diodes are mounted on a 6-Axis Locking Kinematic Mount (Thorlabs K6XS), which is placed on a piezo-actuated stage (Piezosystem Jena Tritor 100) for edge coupling to the photonic chip. The laser diodes are driven by a power supply (Keithley 2220G-30-1) through an electrostatic discharge (ESD) protection and strain relief cable (Thorlabs SR9HA). The microheaters on the chip are driven by two separate sources (Keithley 2220G-30-1 for the bus waveguide phase shifter and Keithley 2400 for the resonator phase shifter) through a three-pin electrical probe. Different cleaved optical fibers from Thorlabs are used to collect the chip output for each spectral range (SM450 for deep blue, blue and green; SM600 for red; and 780HP for near-IR). The fiber-coupled light is sent to an optical spectrum analyzer (Ando AQ6314A) for spectral measurements and to a wavemeter (Advantest Q8326 for the near-IR and HighFinesse WSU-10 for the other spectral ranges) for precise wavelength measurements of the self-injection locked lasers.

# References

1. Niffenegger, R. J. *et al.* Integrated multi-wavelength control of an ion qubit. *Nature* **586**, 538–542 (2020).

2. Moody, G. *et al.* Roadmap on integrated quantum photonics. *J. Phys. Photonics* (2021) doi:10.1088/2515-7647/ac1ef4.

3. Shin, M. C. *et al.* Chip-scale blue light phased array. *Opt. Lett.* **45**, 1934–1937 (2020).

4. Mohanty, A. *et al.* Reconfigurable nanophotonic silicon probes for sub-millisecond deep-brain optical stimulation. *Nat. Biomed. Eng.* **4**, 223–231 (2020).

5. Lanzio, V. *et al.* Small footprint optoelectrodes using ring resonators for passive light localization. *Microsyst. Nanoeng.* **7**, 1–14 (2021).

6. Shamim, M. H. M. *et al.* Investigation of Self-Injection Locked Visible Laser Diodes for High Bit-Rate Visible Light Communication. *IEEE Photonics J.* **10**, 1–11 (2018).

7. Donvalkar, P. S., Savchenkov, A. & Matsko, A. Self-injection locked blue laser. *J. Opt.* **20**, 045801 (2018).

8. Lai, Y.-H. *et al.* 780 nm narrow-linewidth self-injection-locked WGM lasers. in *Laser Resonators, Microresonators, and Beam Control XXII* vol. 11266 112660O (International Society for Optics and Photonics, 2020).

9. Savchenkov, A. A. *et al.* Application of a self-injection locked cyan laser for Barium ion cooling and spectroscopy. *Sci. Rep.* **10**, 16494 (2020).

10. Payne, F. P. & Lacey, J. P. R. A theoretical analysis of scattering loss from planar optical waveguides. *Opt. Quantum Electron.* **26**, 977–986 (1994).

11. Gil-Molina, A. *et al.* Robust Hybrid III-V/Si3N4 Laser with kHz-Linewidth and GHz-Pulling Range. in *Conference on Lasers and Electro-Optics (2020), paper STu3M.4* STu3M.4 (Optical Society of America, 2020). doi:10.1364/CLEO_SI.2020.STu3M.4.

12. Franken, C. A. A. *et al.* A hybrid-integrated diode laser for the visible spectral range. *ArXiv201204563 Phys.* (2020).

13. Stern, B., Ji, X., Okawachi, Y., Gaeta, A. L. & Lipson, M. Battery-operated integrated frequency comb generator. *Nature* **562**, 401–405 (2018).

14. Corato-Zanarella, M. *et al.* Overcoming the Trade-Off Between Loss and Dispersion in Microresonators. in *Conference on Lasers and Electro-Optics (2020), paper STh1J.1* STh1J.1 (Optical Society of America, 2020). doi:10.1364/CLEO_SI.2020.STh1J.1.

15. Ji, X. *et al.* Ultra-low-loss on-chip resonators with sub-milliwatt parametric oscillation threshold. *Optica* **4**, 619–624 (2017).

16. Kondratiev, N. M. *et al.* Self-injection locking of a laser diode to a high-Q WGM microresonator. *Opt. Express* **25**, 28167–28178 (2017).

17. Galiev, R. R. *et al.* Spectrum collapse, narrow linewidth, and Bogatov effect in diode lasers locked to high-Q optical microresonators. *Opt. Express* **26**, 30509–30522 (2018).

18. Liang, W. *et al.* Ultralow noise miniature external cavity semiconductor laser. *Nat. Commun.* **6**, 7371 (2015).

19. Stéphan, G. M., Tam, T. T., Blin, S., Besnard, P. & Têtu, M. Laser line shape and spectral density of frequency noise. *Phys. Rev. A* **71**, 043809 (2005).

20. Zhang, Z. & Yariv, A. A General Relation Between Frequency Noise and Lineshape of Laser Light. *IEEE J. Quantum Electron.* **56**, 1–5 (2020).

21. Petermann, K. *Laser Diode Modulation and Noise*. (Springer Netherlands, 1988). doi:10.1007/978-94-009-2907-4.

## Acknowledgements

This work was supported as part of the Novel Chip-Based Nonlinear Photonic Sources from the Visible to Mid-Infrared funded by the Army Research Office (ARO) under award no. W911NF2110286. The photonic chips fabrication was done in part at the City University of New York Advanced Science Research Center NanoFabrication Facility, in part at the Columbia Nano Initiative (CNI) Shared Lab Facilities at Columbia University, and in part at the Cornell NanoScale Facility, a member of the National Nanotechnology Coordinated Infrastructure (NNCI), which is supported by the National Science Foundation (Grant NNCI-2025233). M.C.S is supported by the Facebook Fellowship Award. M.C.Z. thanks Natalie Janosik for her support and helpful discussions.

## Author contributions

M.C.Z. conceived the chip-scale visible lasers research project. M.C.Z., A.G.M. and A.M. had the initial discussions that shaped the first steps and goals of the project. M.C.Z. designed the photonic devices and experiments with helpful suggestions from A.G.M.. M.C.Z., X.J. and M.C.S. fabricated the photonic chips. M.C.Z. and A.M. performed experiments on some preliminary devices. M.C.Z. performed the experimental measurements and analyzed the results, with crucial suggestions from A.G.M. along the way. M.C.Z. prepared the manuscript. X.J., A.G.M., M.C.S., A.M. and M.L. edited the manuscript. M.L. supervised the project.